\documentclass{sig-alternate}
\usepackage{color} 
\usepackage{booktabs} 
\usepackage{algorithm2e}
\usepackage{amsmath}
\usepackage{amssymb}
\usepackage{url}
\usepackage{subfig}

\newdef{definition}{Definition}
\newtheorem{remark}{Remark}
\begin{document}

\title{Network Sampling Based on NN Representatives} 

\numberofauthors{4} 
\author{
%
\alignauthor 
Milos Kudelka\\
       \affaddr{FEI, VSB - Technical University of Ostrava}\\
       \affaddr{17. listopadu 15, 708 33}\\
       \affaddr{Ostrava - Poruba, Czech Republic}\\
       \email{milos.kudelka@vsb.cz}
\alignauthor
Sarka Zehnalova\\ 
       \affaddr{FEI, VSB - Technical University of Ostrava}\\
       \affaddr{17. listopadu 15, 708 33}\\
       \affaddr{Ostrava - Poruba, Czech Republic}\\
       \email{sarka.zehnalova.st@vsb.cz}
\alignauthor
Jan Platos\\
       \affaddr{FEI, VSB - Technical University of Ostrava}\\
       \affaddr{17. listopadu 15, 708 33}\\
       \affaddr{Ostrava - Poruba, Czech Republic}\\
       \email{jan.platos@vsb.cz}
}
	   
\date{6 January 2014}

\maketitle
\begin{abstract}
The amount of large-scale real data around us increase in size very quickly and so does the necessity to reduce its size by obtaining a representative sample. Such sample allows us to use a great variety of analytical methods, whose direct application on original data would be infeasible. There are many methods used for different purposes and with different results. In this paper we outline a simple and straightforward approach based on analyzing the nearest neighbors (NN) that is generally applicable. This feature is illustrated on experiments with weighted networks and vector data. The properties of the representative sample show that the presented approach maintains very well internal data structures (e.g. clusters and density). Key technical parameters of the approach is low complexity and high scalability. This allows the application of this approach to the area of big data.
\end{abstract}

{ - Information filtering}

\section{Introduction}
In the area of big data, sampling may help facilitate knowledge discovery from large-scale datasets. Using analytical methods it is possible to find patterns and regularities in a sample which is significanly smaller than the original dataset. To be able to validate observed patterns on the original data, it is necessary to have a representative sample which retains certain statistical properties. \\
In the area of complex networks those are topological properties of the network, i.e. degree, clustering coefficient, and eigenvalues, which are used to evaluate the 'goodness' of the sample. Those properties are traditionally defined primarily for unweighted networks, as in \cite{leskovec2006sampling}.
In the area of large datasets with vector data it is usually the assessment of the extent to which the sample maintains clusters and their density \cite{kollios2003efficient, palmer2000density}. Generally, the approaches can be divided into two groups; unbiased (uniform random sampling) and biased.
Approach described in this paper is one of the biased methods and is based on the selection of representatives, i.e. objects that in their surroundings play a more important role than others. In the case of networks it means the selection of nodes into a sample after the assessment of their representativeness. Simply stated, by representativeness we understand the extent to which a node is the nearest neighbor of other nodes in the network.\\
Our method has five important aspects that will be subsequently described in the paper. The first is determinism. A representative sample is uniquely determined by the parameters specified in the method. The second aspect is based on the fact that a key feature used to assess the representativeness is the similarity. The consequence is that the method is applicable to the weighted networks in which the weight of the edges can be understood as the similarity of the nodes. If we accept the non-symmetry of similarity, then the third aspect is the applicability of the method on directed weighted networks. However, in this paper we work with the assumption that similarity is symmetric. The assumption of symmetry leads to the fourth aspect which is locality. A representative sample can be extracted from certain parts of the network (e.g. around the selected node) without the need to analyze the entire network. The method works only with the surroundings of the examined nodes. The locally extracted sample is then a subset of a representative sample of the entire network. That implies the scalability of the presented algorithm. The last, fifth aspect, is the applicability of the approach on other than the network data. If we have data records represented by points (vectors) in an n-dimensional space, then the distance between the points can be interpreted as dissimilarity. By converting dissimilarity into similarity, the problem of obtaining a sample from an n-dimensional dataset in the vector space can be modified to the problem of finding a sample of the network. This aspect will be  analyzed in more detail in Section \ref{sec:exp}.\\
We illustrate the usage of our method on experiments with small network and 2-dimensional dataset as well as on two large real world datasets. The first is the weighted co-authorship network with more than 300,000 nodes based on the DBLP database. The second is a list of all address points in the Czech Republic, 2.7 million points in total.


\section{Related work}
\label{sec:rel}
Sampling has already been applied to various types of data in many areas. For large n-dimensional datasets sampling methods have been developed in order optimize data mining tasks \cite{kerdprasop2005weighted,kollios2003efficient,nanopoulos2002efficient,palmer2000density}, such as clustering or outlier detection.\\
In the area of networks, sampling goals range from speeding up simulations \cite{krishnamurthy2007sampling} to refined visualization \cite{rafiei2005effectively}. The evaluation of sampling algorithms is related to the sampling goals \cite{ahmed2010reconsidering, rasti2008evaluating}. In the area of large-scale social networks a work of Leskovec et al. \cite{leskovec2006sampling} compared state-of-the-art sampling algorithms and defined representative subgraph sampling based on several topological properties. In \cite{ hubler2008metropolis} proposed Metropolis subgraph sampling, where the measuring of graph properties is included as a sampling step. On sampling community structure focused in \cite{maiya2010sampling}, where the goal was to preserve and infer community affiliation of nodes in the network.\\
All of the above mentioned methods produce a random sample from an original dataset, while in our approach sampling is deterministic. Therefore, the resulting representative sample is determined by the configuration of the algorithm.  

\section{Proposed method} 
\label{sec:met}
In this section we define the key concepts precisely and propose a general sampling approach based on representatives.
\subsection{Notations and Definitions}
Let $\mathbf{D}$ be the dataset. For a dataset consisting of vector data, sample $ \mathbf{S}$ is a subset of those vectors. For a network or graph $G = (V, E)$, $\mathbf{S}$ is a sample of nodes where $S \subset V$. \\


\noindent
{\em Object} is a basic entity from a dataset $D$. \\

\noindent
 {\em Similarity} is a relationship between two {\em objects}. It is a function $s: \mathbf{D} \times \mathbf{D} \rightarrow \mathbb{R}_{0}^{+}$. \\
 

\noindent
{\em Proximity} is a {\em similarity} greater than a specific threshold. It is a function $p: \mathbf{D} \times \mathbf{D} \rightarrow \{0,1\}$. {\em Object} $o$ is {\em close} to {\em object} $o'$ if $p(o,o') = 1$. \\

\noindent
Large similarity is equivalent to a small distance. \\

\noindent
{\em Neighbor} is an {\em object} {\em close} to an examined {\em object}.\\

\noindent
{\em Neighborhood} is a set of all {\em neighbors} of an examined {\em object}, $\varepsilon(o)=\{o' \in D : p(o,o')=1\}$. \\

\noindent
{\em Nearest neighbor} is a {\em neighbor} that is closest to an examined {\em object}. An {\em object} can have more than one {\em nearest neighbor}. \\

\noindent
The importance of an object is determined by its relationship to its surroundings. Generally, significant is the number of neighborhoods in which the object is present and how many times it is the nearest neighbor. \\

\noindent
{\em Proximity degree} is the number of {\em objects} that are {\em neighbors} of an examined {\em object} and also the number of {\em neighborhoods} that contain the examined {\em object}. It is a function $d: \mathbf{D} \rightarrow \mathbb{N}_{0}$. \\

\noindent
{\em Proximity rank} is the number of {\em objects} to which an examined {\em object} is a {\em nearest neighbor}. It is a function $k: \mathbf{D} \rightarrow \mathbb{N}_{0}$. \\

\noindent
{\em Representativeness} is the property that expresses the extent to which the examined object can represent other objects in the dataset. {\em Representativeness} should be based mainly on the {\em proximity rank}, but may also reflect the {\em proximity degree}. It is a function $r: \mathbf{D} \rightarrow \mathbb{R}_{0}^{+}$. \\

\noindent
{\em Representative} is an {\em object} whose {\em representativeness} is equal to or greater than a constant (or otherwise defined) threshold.
A function $f: \mathbf{D} \rightarrow \{0,1\}$ describes that {\em objects} $o$ is a {\em representative} of a set $D$. \\

\noindent
On the basis of the representatives, it is possible to define a sample of objects that represent the entire dataset. \\

\noindent
{\em Representative sample} is a set of all {\em representatives} of a dataset. It is a set $S = \{o \in D : f(o)=1\}$ which is defined by $(\mathbf{D},s,p,r,f)$. 

\begin{remark}
The functions of similarity and proximity do not have to be symmetric.
\end{remark}

\subsection{Sampling algorithm}
\label{sec:alg}
Our aproach to obtaining a representative sample of a dataset is inpired by the method of finding nearest neighbors. The algorithm is based on the idea, that objects, which are the nearest neighbors of other objects, are the important ones in a dataset. It is a local oriented algorithm, which reduces the given dataset to its sample. 


\begin{algorithm}
\SetKwInOut{Input}{input}\SetKwInOut{Output}{output}
\Input{dataset $D$}
\Output{sample $S$}
\SetKwIF{If}{ElseIf}{Else}{if}{then}{else if}{else}{endif}
\SetAlgoLined
choose a function for $similarity$, $proximity$, $representativeness$ and $representative$\\

\ForEach{object $o \in D$}
{ 
 find neighborhood $\varepsilon(o)$ and a set of nearest neighbors $NN(o)$\\
 \ForEach{object $x \in \varepsilon(o)$}
 {
    increase its $proximity$ $degree$
 }
 \ForEach{object $y \in NN(o)$}
 {
    increase its $proximity$ $rank$
 }
}

\ForEach{object $o \in D$}
{
 \If{$o$ is a $representative$}
 {
  add $o$ to $S$\\
 }
}
\caption{Sampling algorithm}\label{alg:1}
\end{algorithm}

\section{Experimental evaluation}
\label{sec:exp}
\subsection{Datasets and setup}
We tested our algorithm on the 4 following datasets. Two of them representing weighted networks (Miserables \cite{knuth1993stanford} and DBLP\footnote{\url{http://www.informatik.uni-trier.de/~ley/db/}}, a co-authorship network). And other two representing a 2-dimensional data (Birch3 \cite{zhang1997birch}, synthetic data with 100,000 vectors and Czech republic map, a set of all address points in the Czech Republic). For datasets characteristics see Table \ref{tab:data}, where n = dataset size, e = number of edges, dim = dataset dimensionality.
\begin{table}[htbp]
  \centering
  \caption{Characteristics of Datasets}
    \begin{tabular}{rrrrr}
    \toprule
    \textbf{dataset} & \textbf{n} & \textbf{e} & \textbf{dim} & \textbf{type} \\
    \midrule
    Miserables & 77  & 254 &  -  & weighted network \\
    DBLP & 318,971 & 786,384 &  -  & weighted network \\
	Birch3 & 100,000 &  -  & 2   & vector data \\
    Czech map & 2,740,903 &  -  & 2   & vector data \\
    \bottomrule
    \end{tabular}%
  \label{tab:data}%
\end{table}%

In all experiments we use the function of {\em representativeness} $r(o) = \frac{k(o)}{\log _x d\left(o\right)}$. For $d(o)=0$ we define $r(o)=0$. For $d(o)=1$ we define $r(o)=k(o)$.
Resulting size of the sample depends on two parameters. The first is the logarithm base $x$, the second is the function $f$ that is used to define {\em representativeness}. For our experiments we chose  $f: r(o) \geq 1$. We must also define the functions for the {\em similarity}, {\em proximity}, which is done separately in the experiments. We were looking for representative samples of different sizes for each dataset.  For every obtained sample, we give the summary of its size as a percentage against the original dataset, the base of the logarithm in function $r$ and some other characteristics. 

\subsection{Experiment 1 - Weighted Networks}
We choose the weight of an edge as a similarity function. The proximity is defined as follows; close to an examined node are all of its adjacent nodes. 

\begin{figure*}[ht!]
\minipage{0.24\textwidth}
  \includegraphics[width=\linewidth]{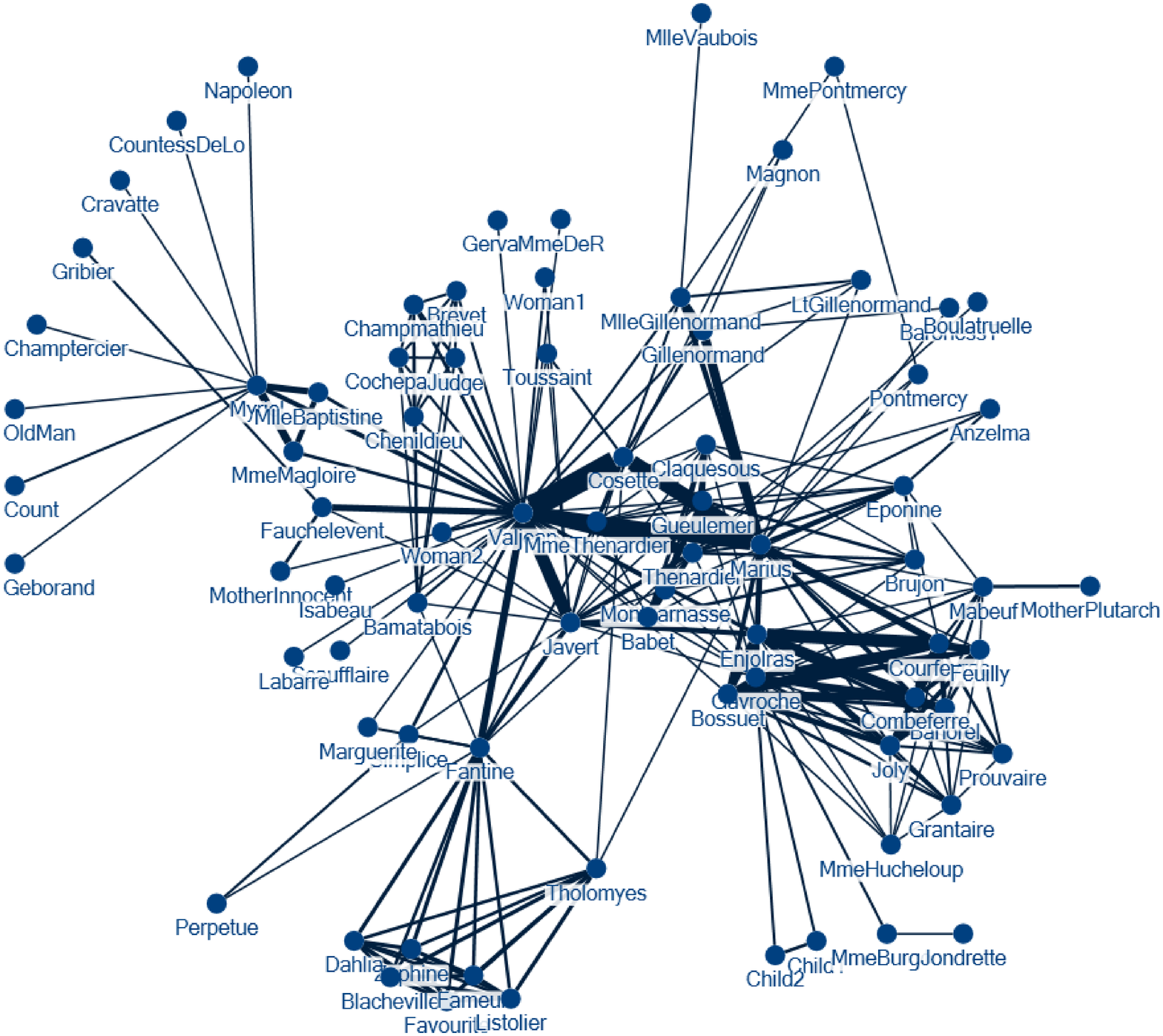}
  \caption{Miserables}\label{fig:mis}
\endminipage\hfill
\minipage{0.24\textwidth}
  \includegraphics[width=\linewidth]{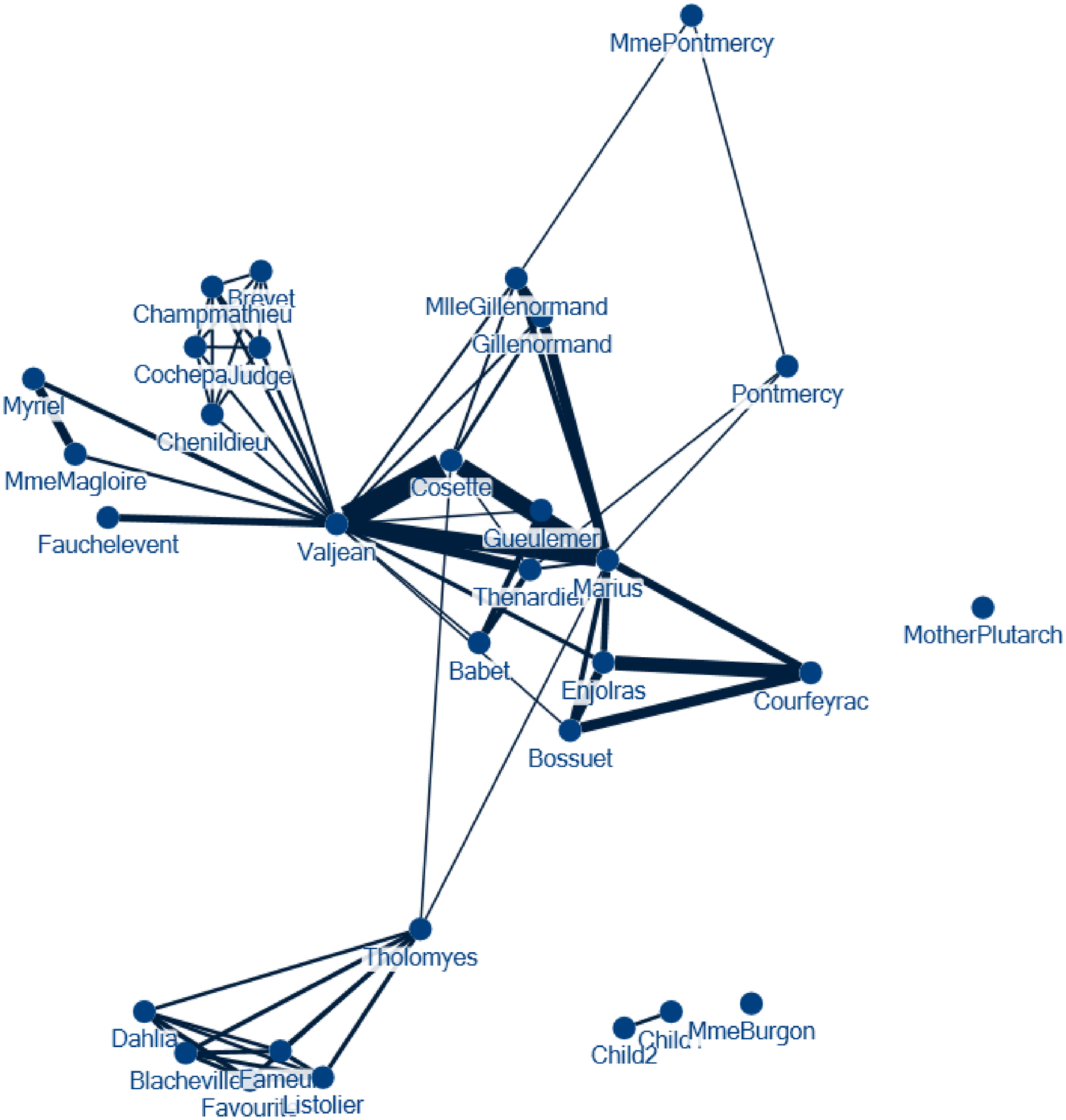}
  \caption{Sample 40\%}\label{fig:mis1}
\endminipage\hfill
\minipage{0.24\textwidth}%
  \includegraphics[width=\linewidth]{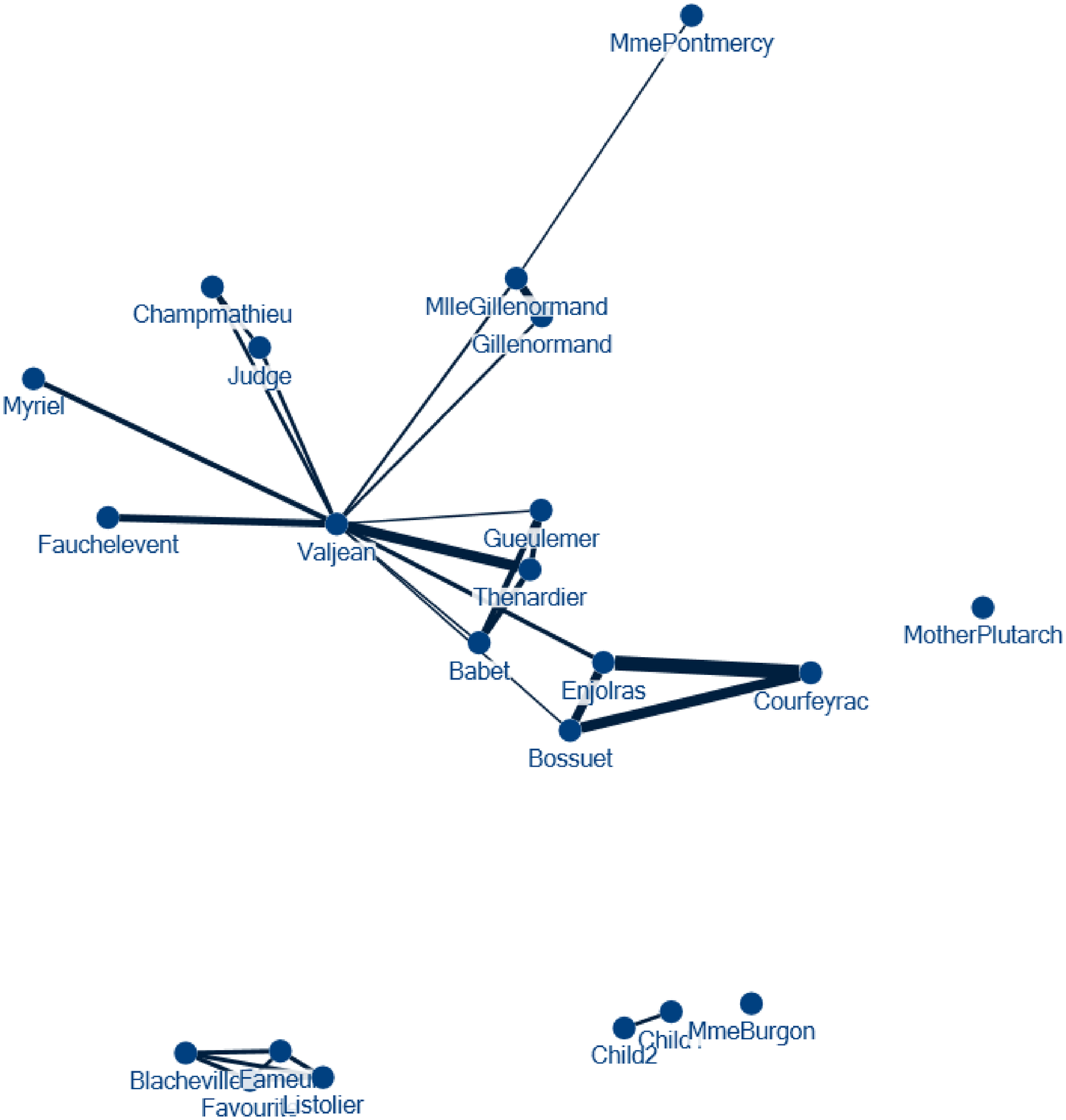}
  \caption{Sample 29\%}\label{fig:mis2}
\endminipage
\minipage{0.24\textwidth}
  \includegraphics[width=\linewidth]{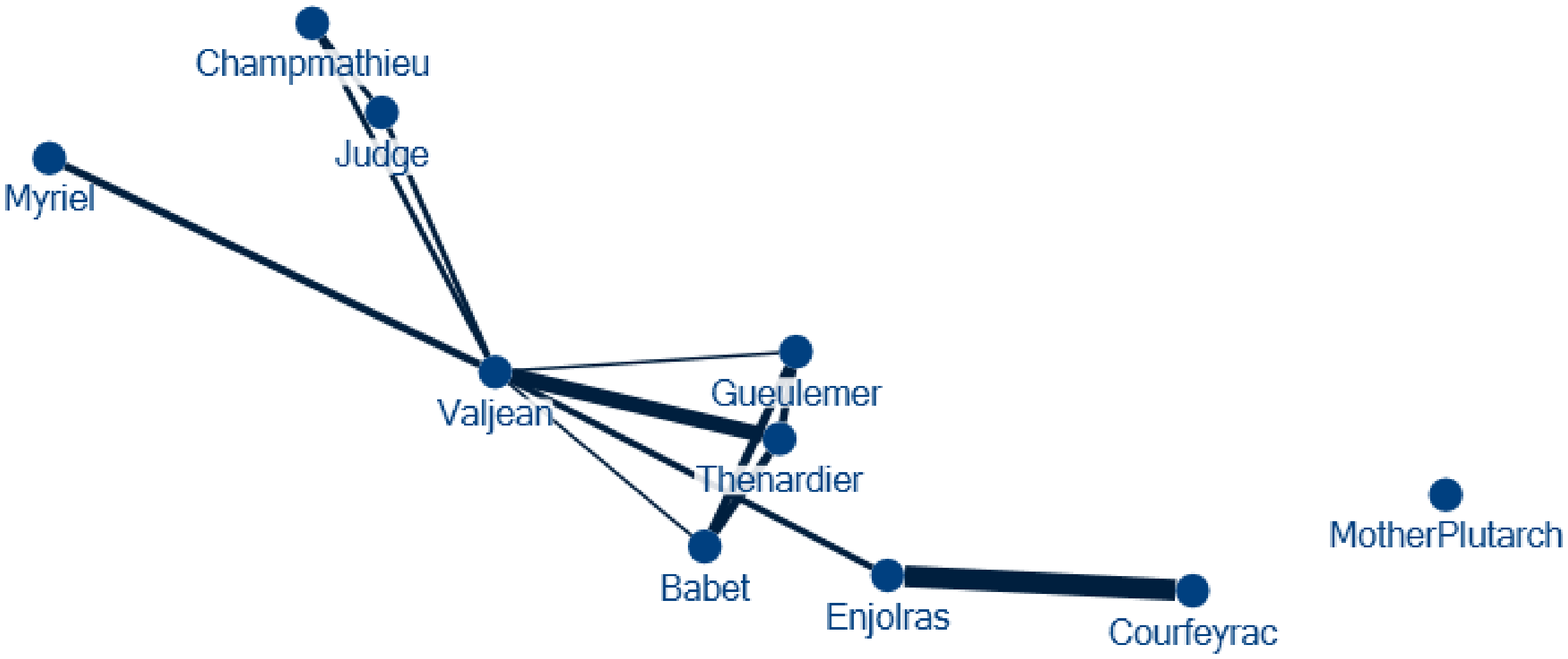}
  \caption{Sample 13\%}\label{fig:mis3}
\endminipage\hfill
\end{figure*}

First dataset is a co-appearance network of characters in the novel Les Miserables in which the edges have weights from 1 to 31, see Figure \ref{fig:mis}. Table \ref{tab:mis}
summarizes the results of the sampling. For the resulting sample networks see Figures \ref{fig:mis1}-\ref{fig:mis3}. 

\begin{table}[htbp]
  \centering
  \caption{Les Miserables Dataset Sampling}
    \begin{tabular}{rrrrr}
    \toprule
    \textbf{Log base} & \textbf{Nodes} & \textbf{Edges} & \textbf{Nodes \%} & \textbf{Edges \%} \\
    \midrule
    -   & 77  & 254 & 100 & 100 \\
    3   & 31  & 67  & 40  & 26 \\
    2   & 22  & 27  & 29  & 11 \\
    1.8 & 10  & 12  & 13  & 5 \\
    \bottomrule
    \end{tabular}%
  \label{tab:mis}%
\end{table}%

Second dataset\footnote{This dataset can be freely downloaded from \url{http://www.forcoa.net/resources/www2014/data_2012_12.zip}} is a co-authorship network contructed from a DBLP dataset. Data were downloaded in april 2013 and preprocessed for the Forcoa.NET system. Edge weights are based on network evolution and forgetting function that takes into account the frequency and regularity of publishing, for details see \cite{kudvelka2012social}.
After the preprocessing a total of 318,971 nodes (active authors) and 786,384 edges (with a maximum weight of 433.459) remained in the network. Table \ref{tab:dblp} summarizes the results of the sampling. Figure \ref{fig:mis} captures co-authors of Christos Faloutsos as a weighted network. For the resulting samples from this part of network see Figures \ref{fig:mis1}-\ref{fig:mis3}.

\begin{table}[htbp]
  \centering
  \caption{DBLP Dataset Sampling}
    \begin{tabular}{rrrrr}
    \toprule
    \textbf{Log base} & \textbf{Nodes} & \textbf{Edges} & \textbf{Nodes \%} & \textbf{Edges \%} \\
    \midrule
    -   & 318,971 & 786,384 & 100 & 100 \\
    2   & 180,694 & 404,624 & 57  & 51 \\
    1.5 & 112,885 & 213,758 & 35  & 27 \\
    1.3 & 37,287 & 67,129 & 12  & 9 \\
    \bottomrule
    \end{tabular}%
  \label{tab:dblp}%
\end{table}%

\begin{figure}[h]
     \centering
     \subfloat[][Node degrees]{\includegraphics[width=0.5\linewidth]{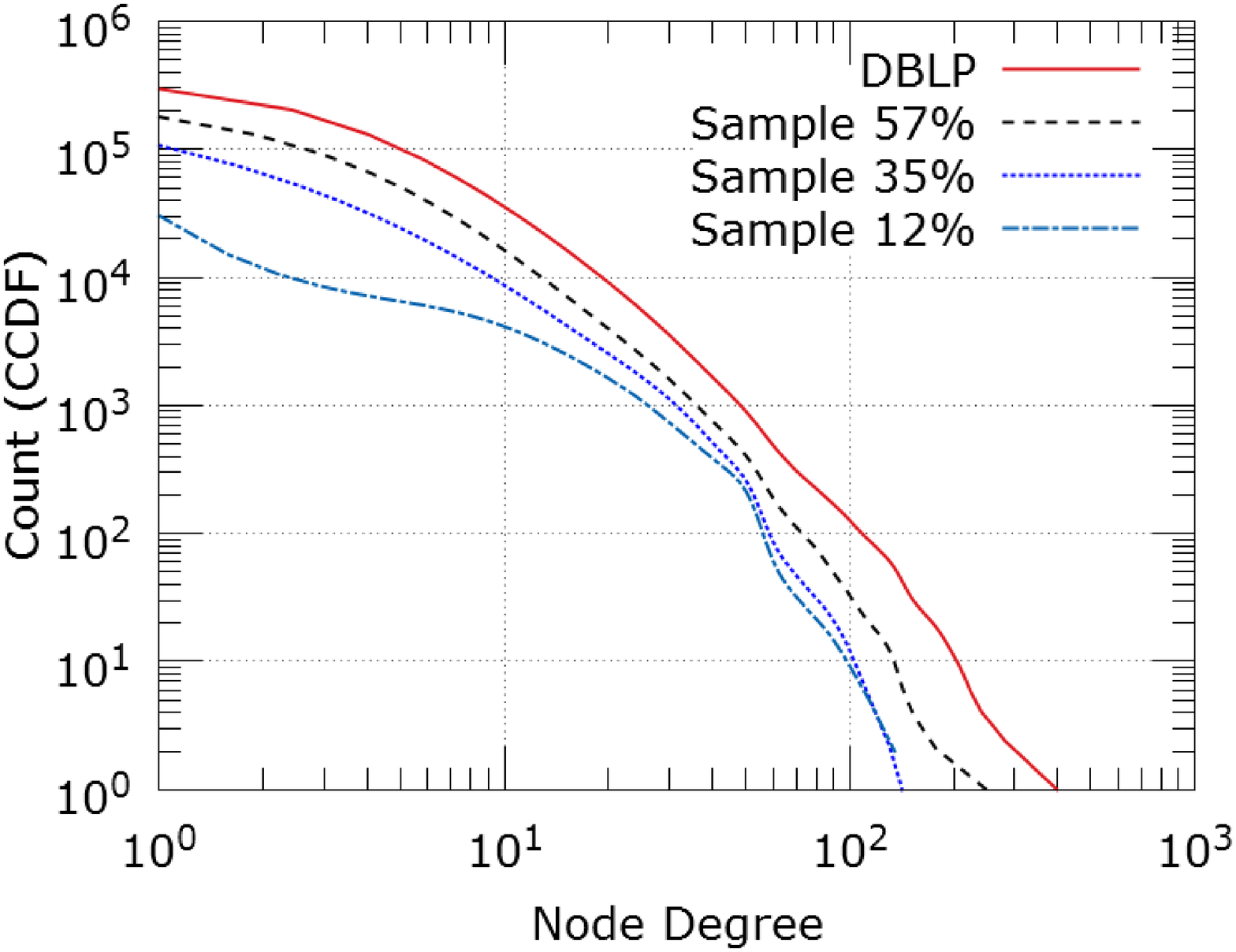}\label{fig:dblp_CD}}
     \subfloat[][Edge weights]{\includegraphics[width=0.5\linewidth]{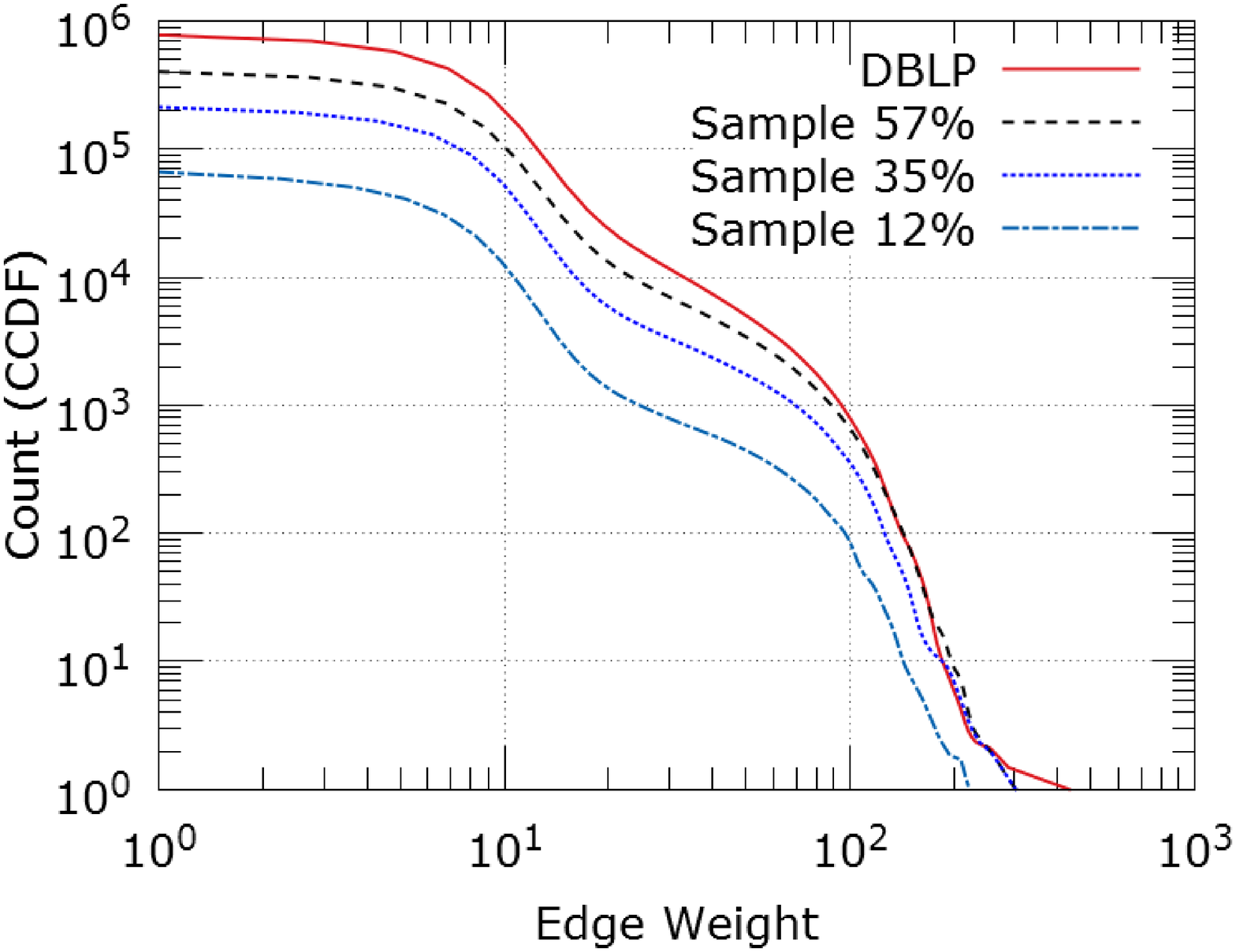}\label{fig:dblp_CW}}
     \caption{DBLP sample metrics}
     \label{fig:dblp_CDCW}
\end{figure}

How some of the topological properties of sampled networks are maintained is depicted on Figure \ref{fig:dblp_CDCW}. The cumulative degree and edge weight distributions clearly copy the distributions of the original network. 


\begin{figure*}[ht]
\minipage{0.24\textwidth}
  \includegraphics[width=\linewidth]{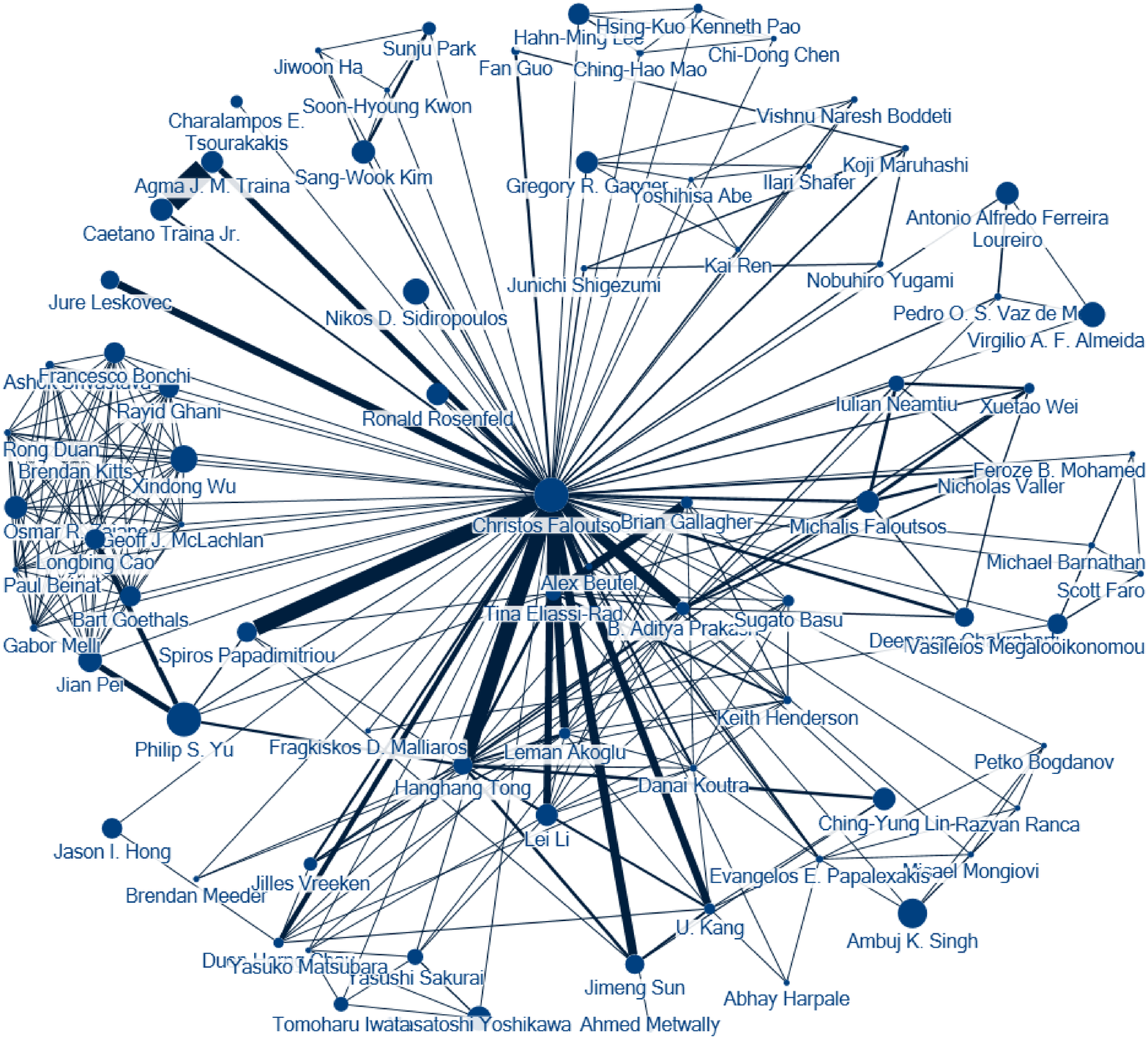}
  \caption{DBLP Faloutsos}\label{fig:fal}
\endminipage\hfill
\minipage{0.24\textwidth}
  \includegraphics[width=\linewidth]{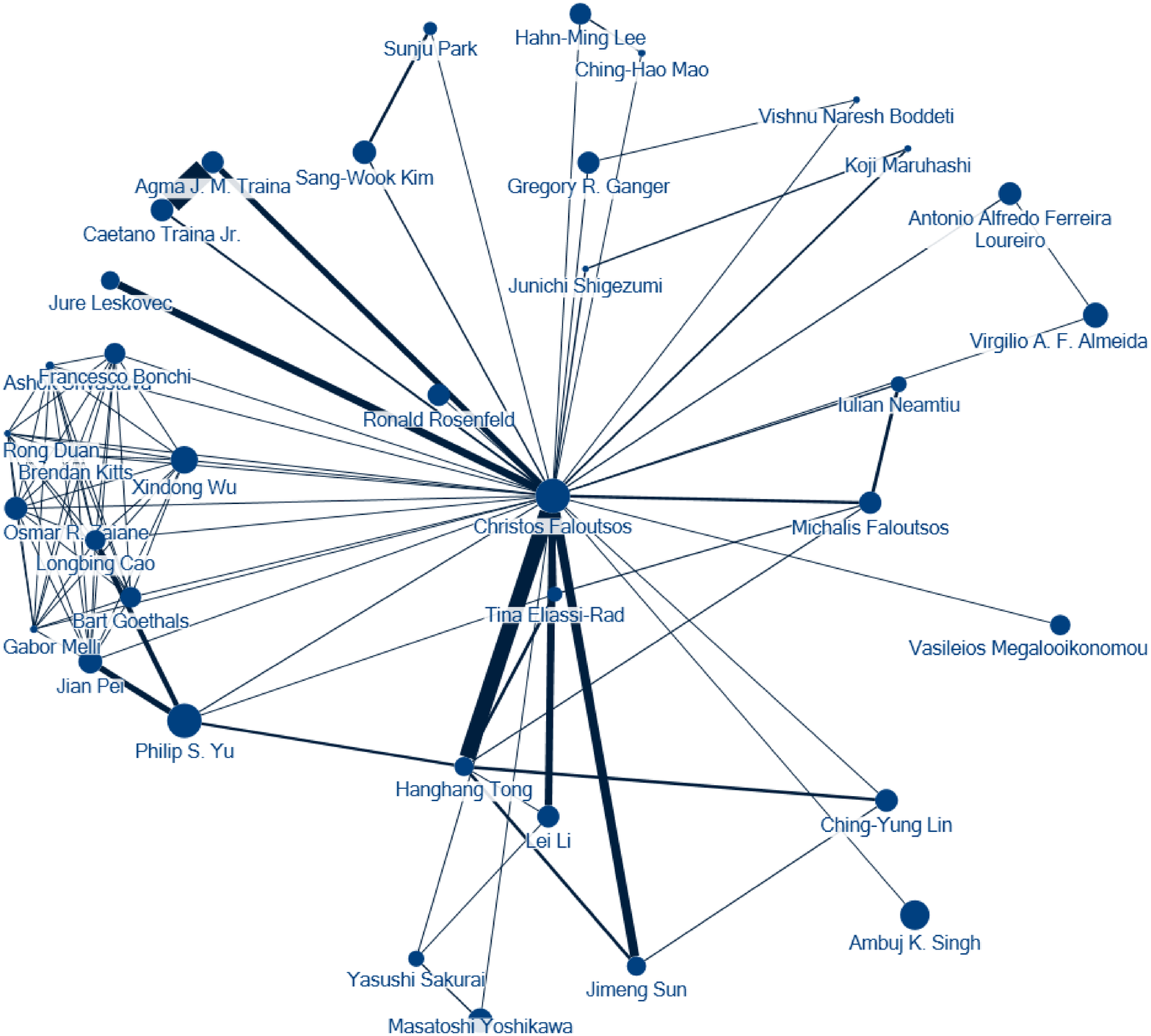}
  \caption{Sample 57\%}\label{fig:fal1}
\endminipage\hfill
\minipage{0.24\textwidth}%
  \includegraphics[width=\linewidth]{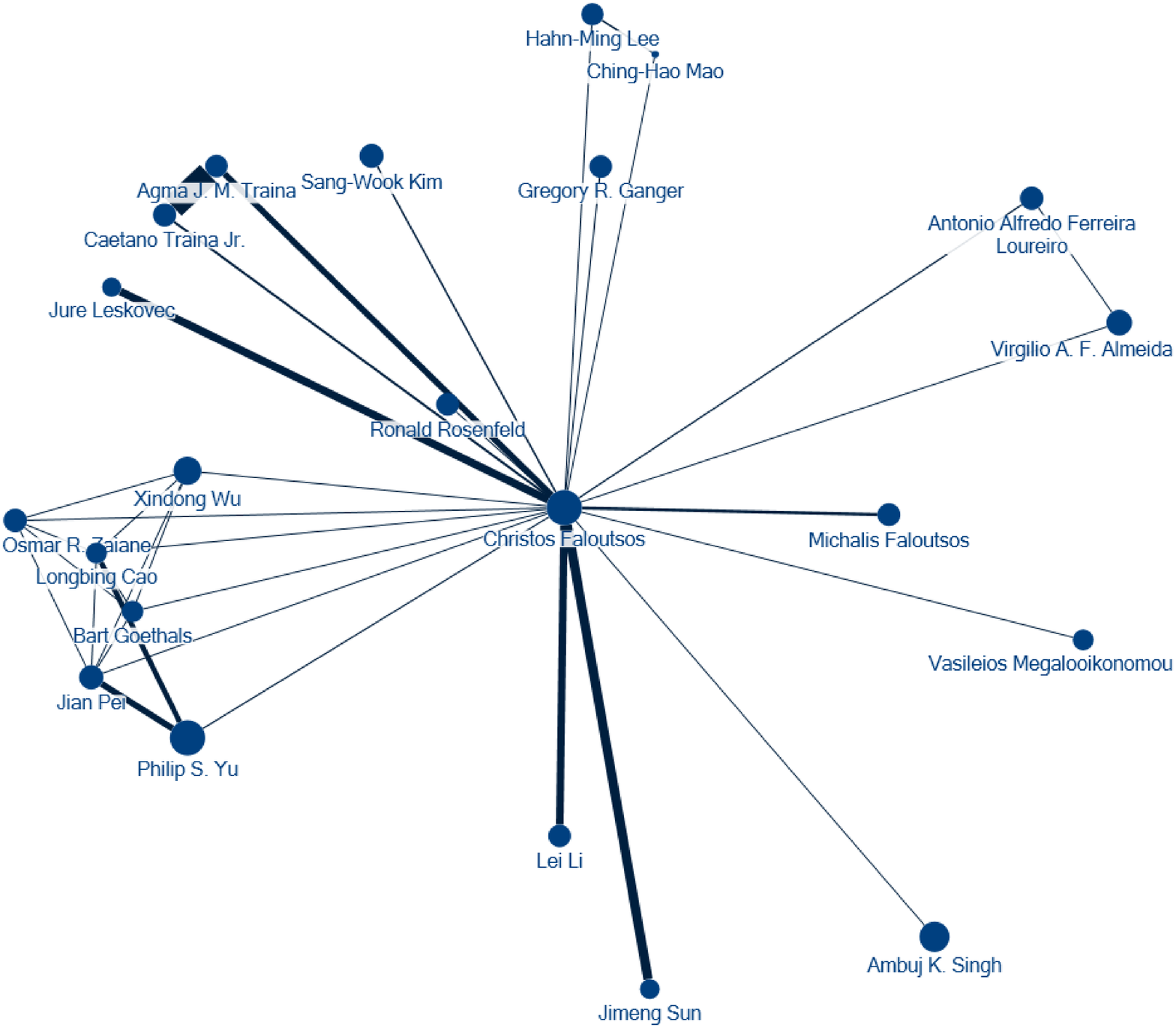}
  \caption{Sample 35\%}\label{fig:fal2}
\endminipage
\minipage{0.24\textwidth}
  \includegraphics[width=\linewidth]{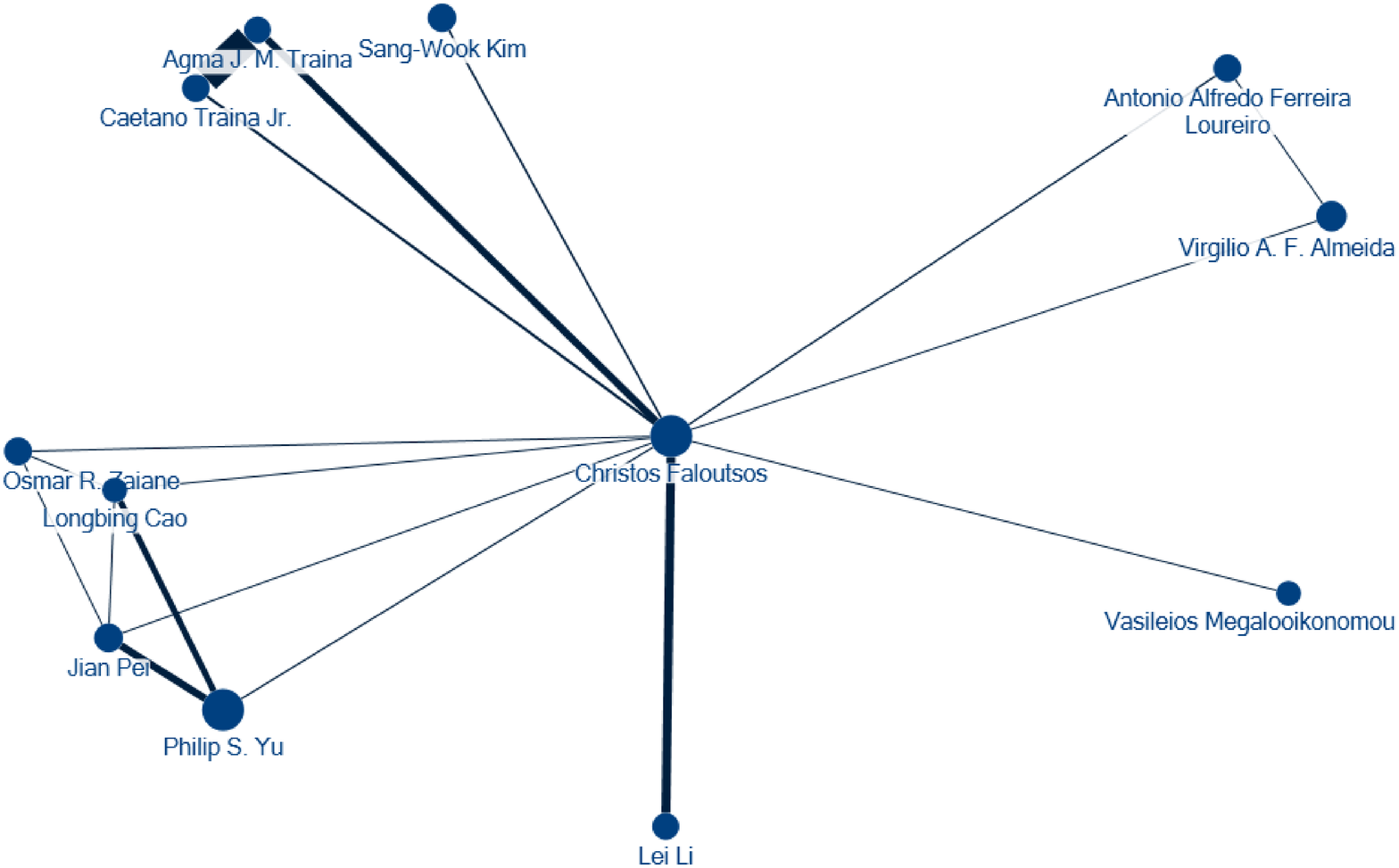}
  \caption{Sample 12\%}\label{fig:fal3}
\endminipage\hfill
\end{figure*}

\subsection{Experiment 2 - Vector Data}
The experiments with vector data are focused mainly on the ability of the suggested algorithm to reduce data while preserving the important features such as clusters. 

We choose the Euclidean distance as a dissimilarity function which is a measure in metric space. The proximity was computed according the predefined maximal distance to the examined object. The neighborhood is defined as a set of objects (vectors) with distance less or equal to a defined radius and the nearest neighbors are objects with the minimal distance to the examined object. The representativeness of an object is defined with the same logarithmic fraction as was used in experiments with network data. 

When we deal with vector data and metric spaces in the way we defined using our dissimilarity function we have two parameters which may be tuned according the expected result - logarithm base and radius. The experiments shows that these two parameters are tied together, because we may produce similar results with larger logarithm base and radius as we produced with smaller base and smaller radius. Even the number of preserved points is similar. Therefore, we experimentally set the logarithm base and then we change the radius in the experiments. 

\begin{figure*}[h]
\minipage{0.24\textwidth}
  \includegraphics[width=\linewidth]{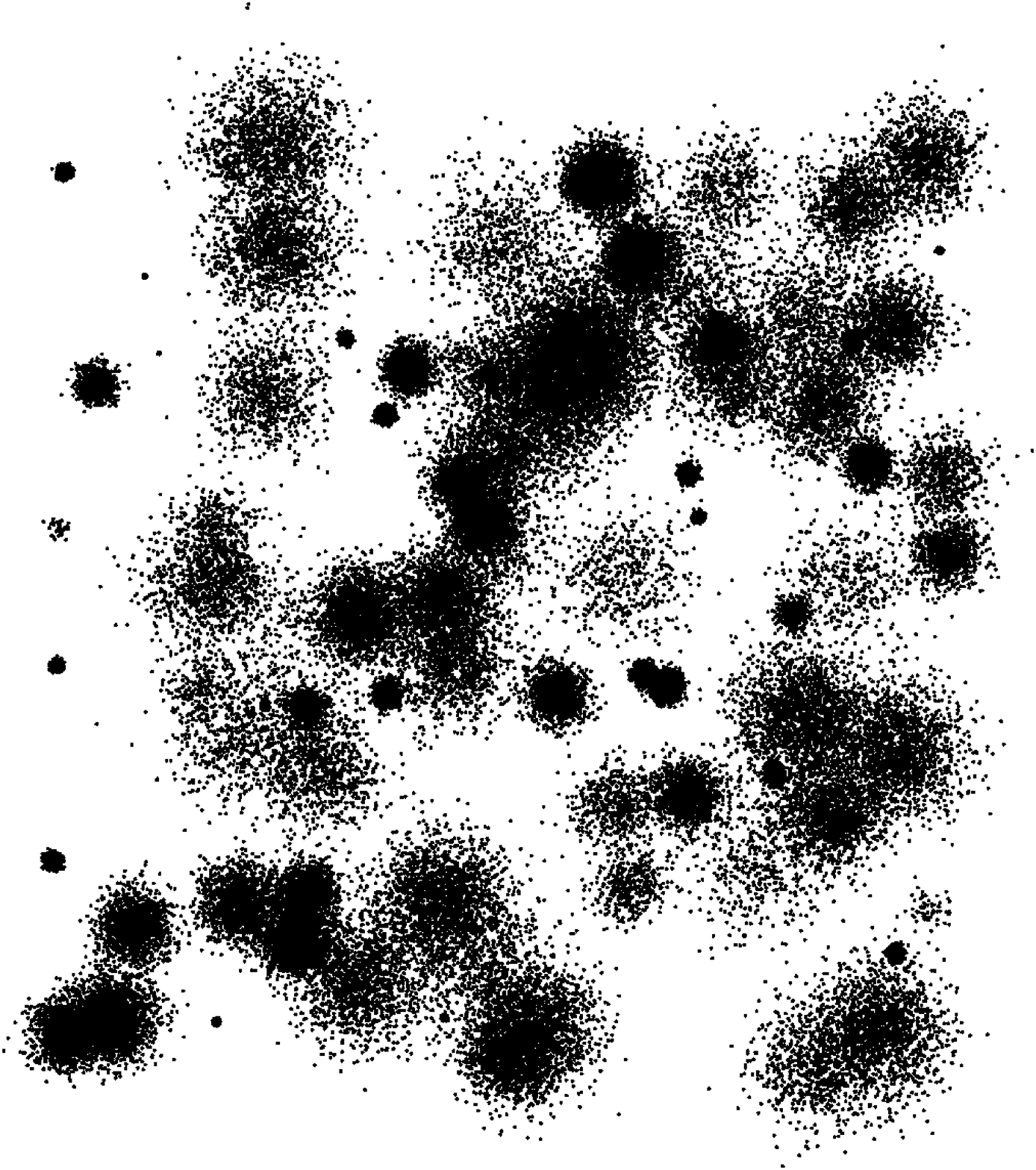}
  \caption{Orig. dataset}
	\label{fig:s1}
\endminipage\hfill
\minipage{0.24\textwidth}
  \includegraphics[width=\linewidth]{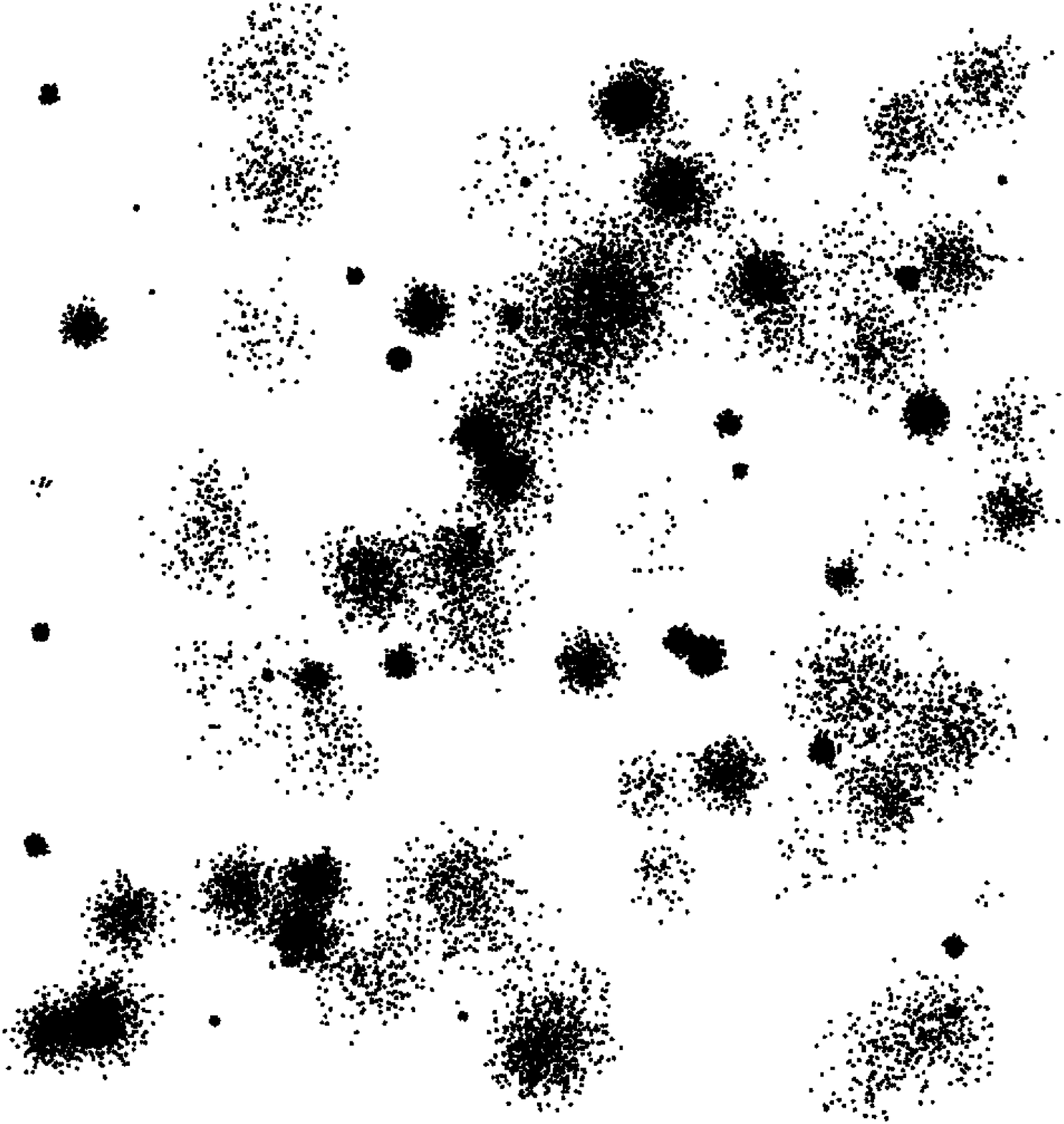}
  \caption{Sample 44\%}
	\label{fig:s1-5000}
\endminipage\hfill
\minipage{0.24\textwidth}
  \includegraphics[width=\linewidth]{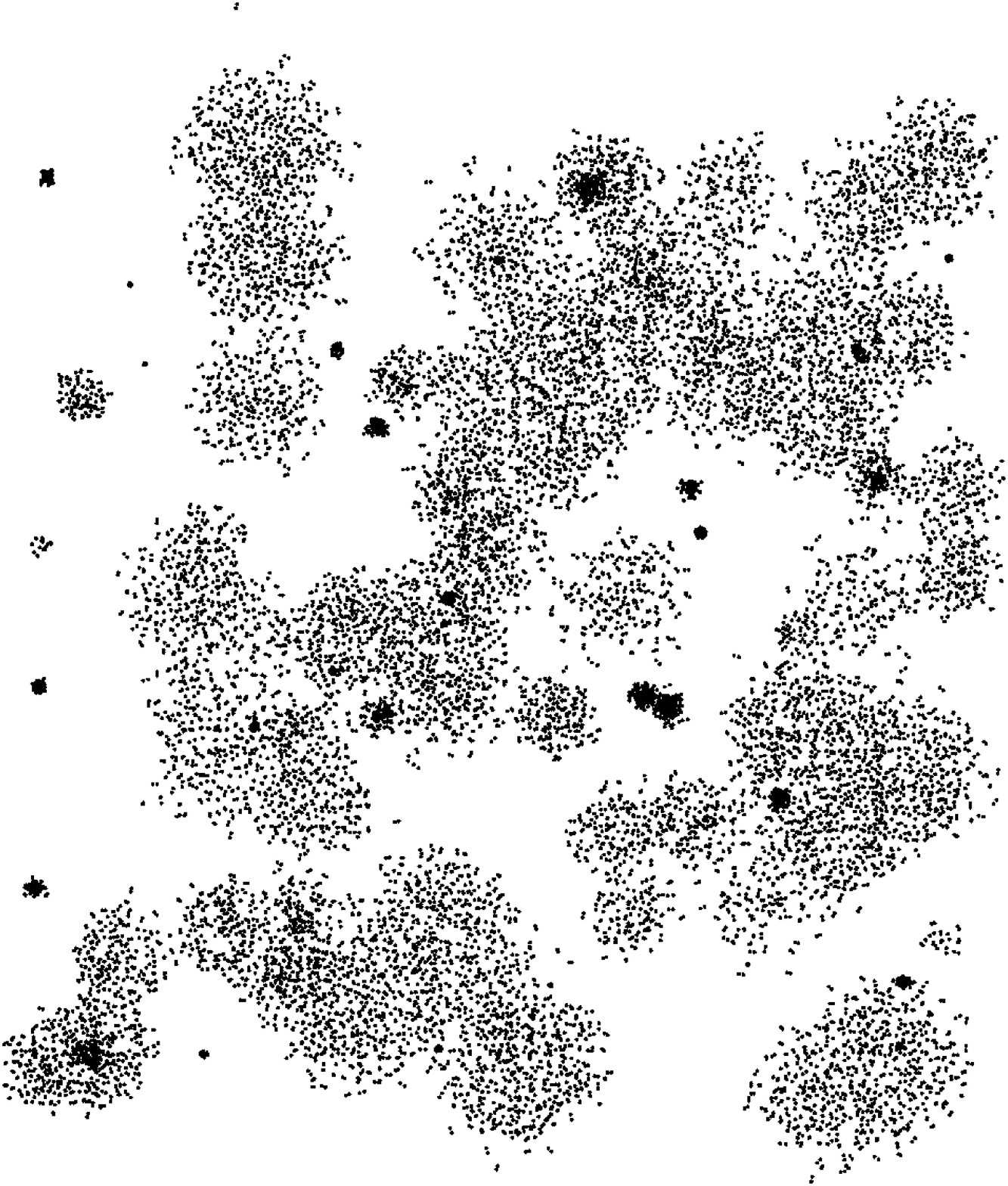}
  \caption{Sample 28\%}
	\label{fig:s1-10000}
\endminipage\hfill
\minipage{0.24\textwidth}
  \includegraphics[width=\linewidth]{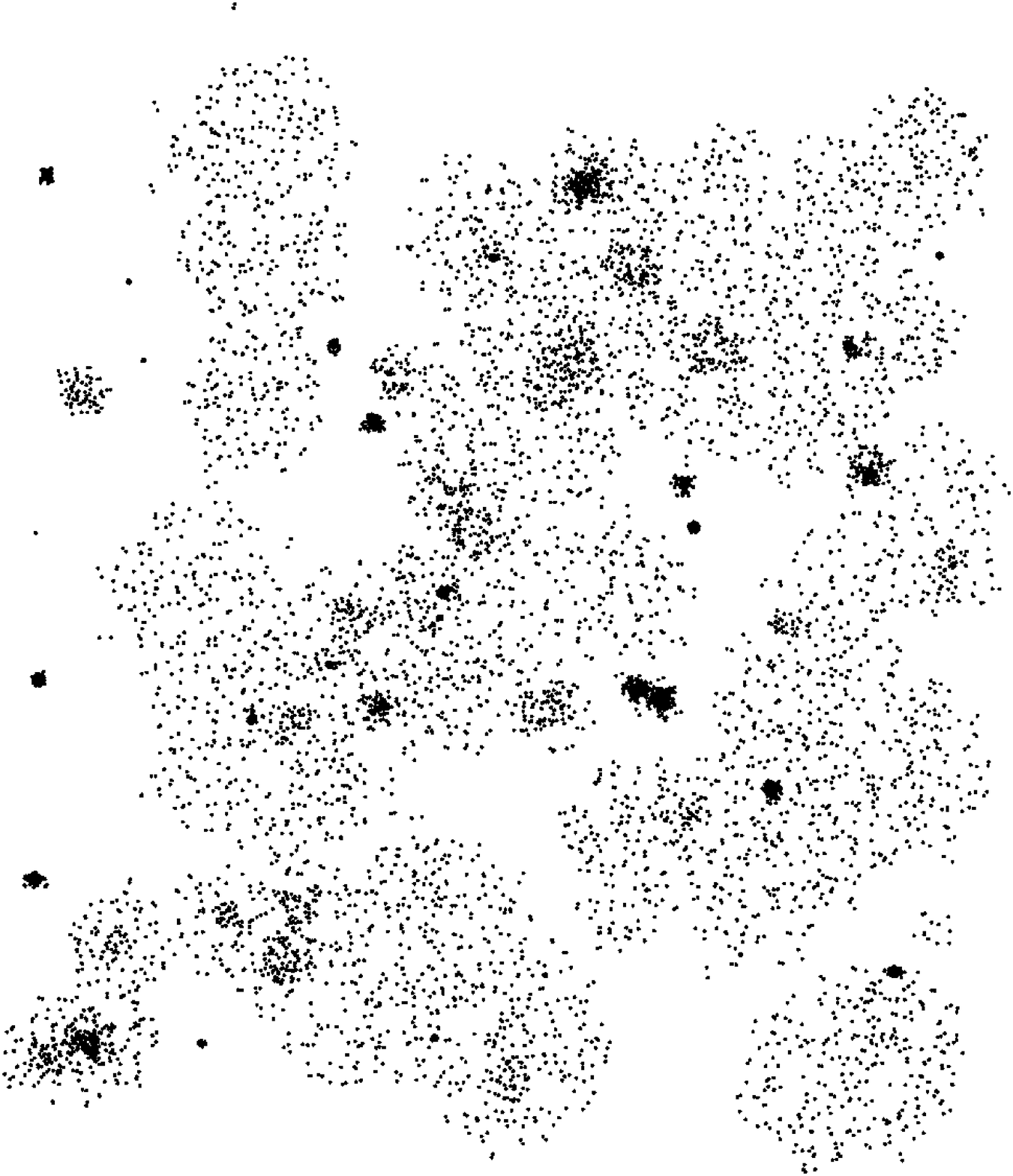}
  \caption{Sample 15\%}
	\label{fig:s1-20000}
\endminipage\hfill
\end{figure*}

The first (synthetic) dataset contains several more or less dense, random sized, clusters in random locations. This dataset has 100,000 points. 
Additionally to the dissimilarity function as an Euclidean distance, we discretize the distance with step 100. We experimentally set the base of the logarithm to 4 and the radius of the neighborhood to 50, 100, and 200 units.
The summary of the experiment is depicted in Table \ref{tab:cluster} and the visualization of the data is on Figures \ref{fig:s1}-\ref{fig:s1-20000}. 

\begin{table}[htbp]
  \centering
  \caption{Birch3 Dataset Sampling}
    \begin{tabular}{rrrr}
    \toprule
    \textbf{Log base} & \textbf{Radius} & \textbf{Points} & \textbf{Points \%} \\
    \midrule
    -   & -  & 100,000 & 100  \\
    4   & 50  & 44,098  & 44   \\
    4   & 100  & 24,745  & 28   \\
    4 & 200  & 14,835  & 15   \\
    \bottomrule
    \end{tabular}%
  \label{tab:cluster}%
\end{table}%

\begin{figure*}[b]
\minipage{0.49\textwidth}
  \includegraphics[width=\linewidth]{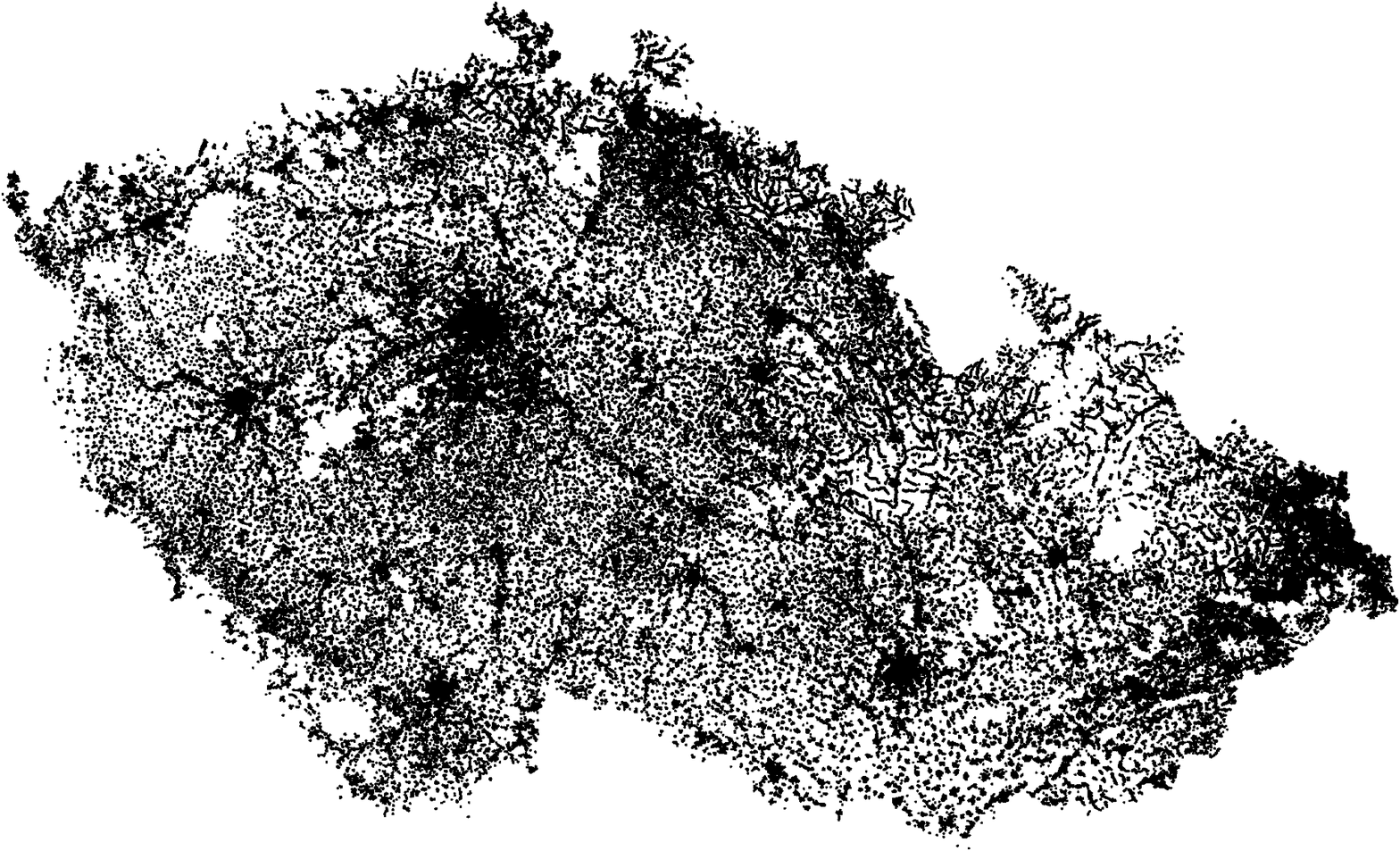}
  \caption{Original dataset}
	\label{fig:address}
\endminipage\hfill
\minipage{0.49\textwidth}
  \includegraphics[width=\linewidth]{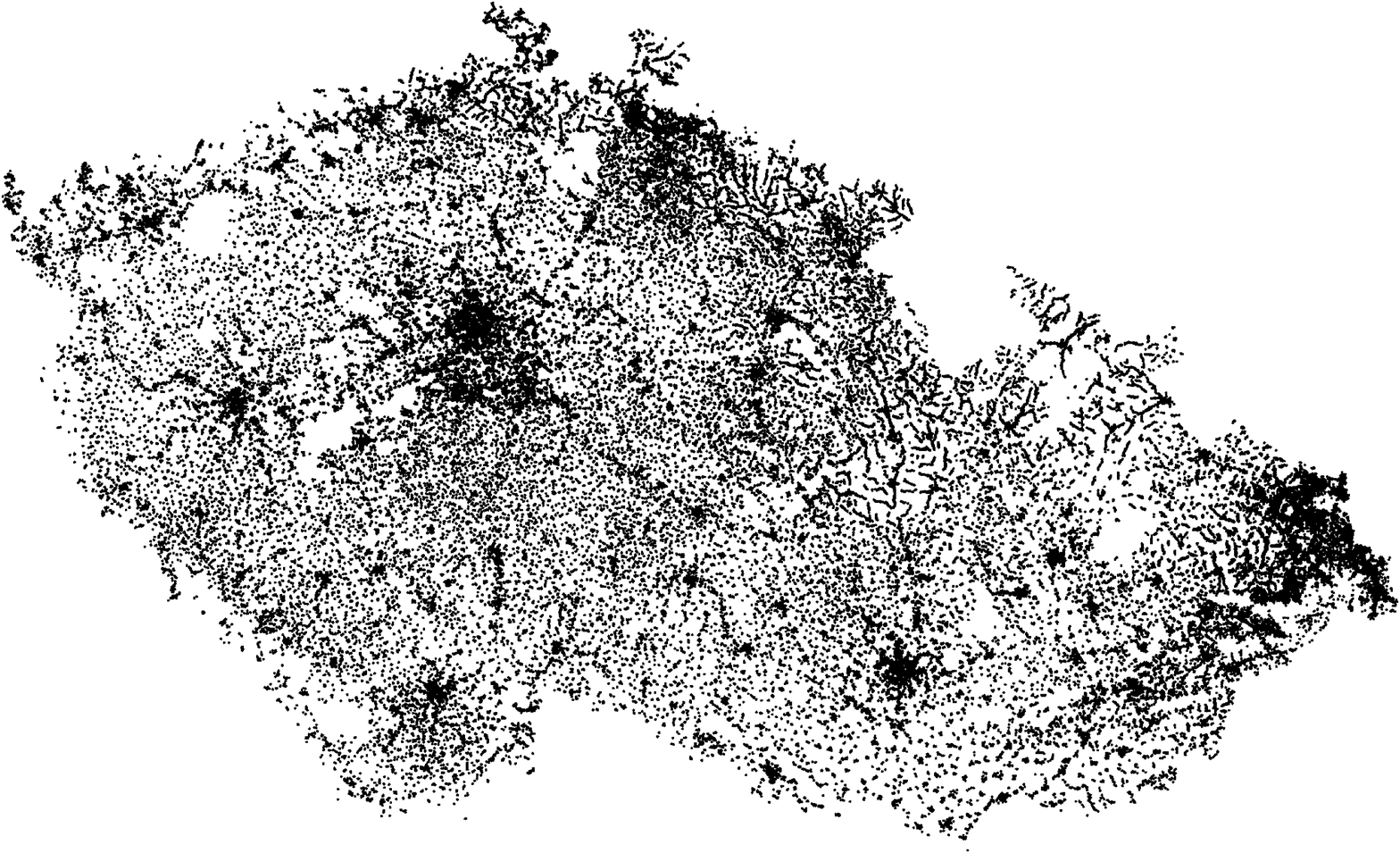}
  \caption{Sample 8\%}
	\label{fig:address050}
\endminipage\hfill
\\[10mm]
\minipage{0.49\textwidth}
  \includegraphics[width=\linewidth]{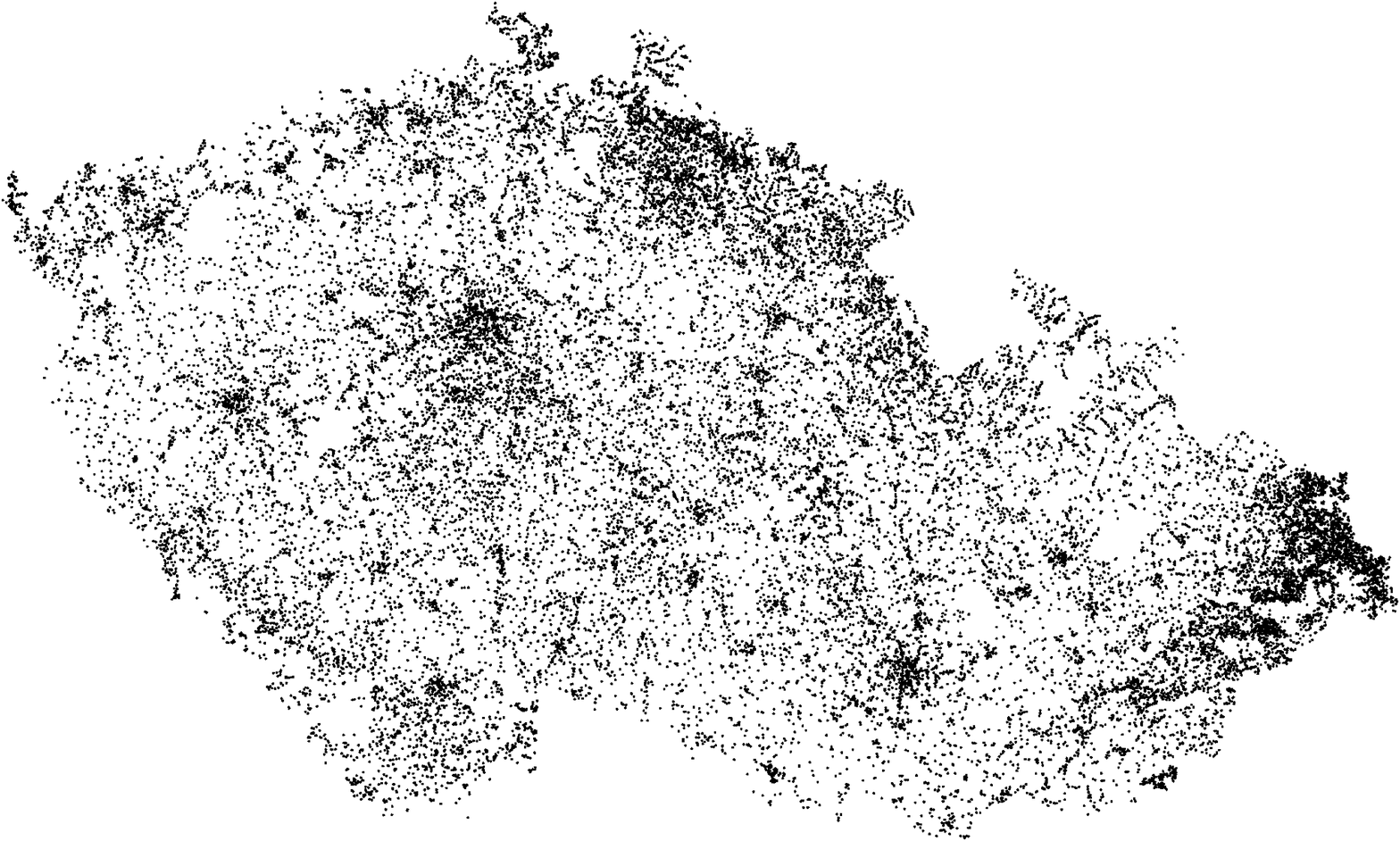}
  \caption{Sample 2\%}
	\label{fig:address100}
\endminipage\hfill
\minipage{0.49\textwidth}
  \includegraphics[width=\linewidth]{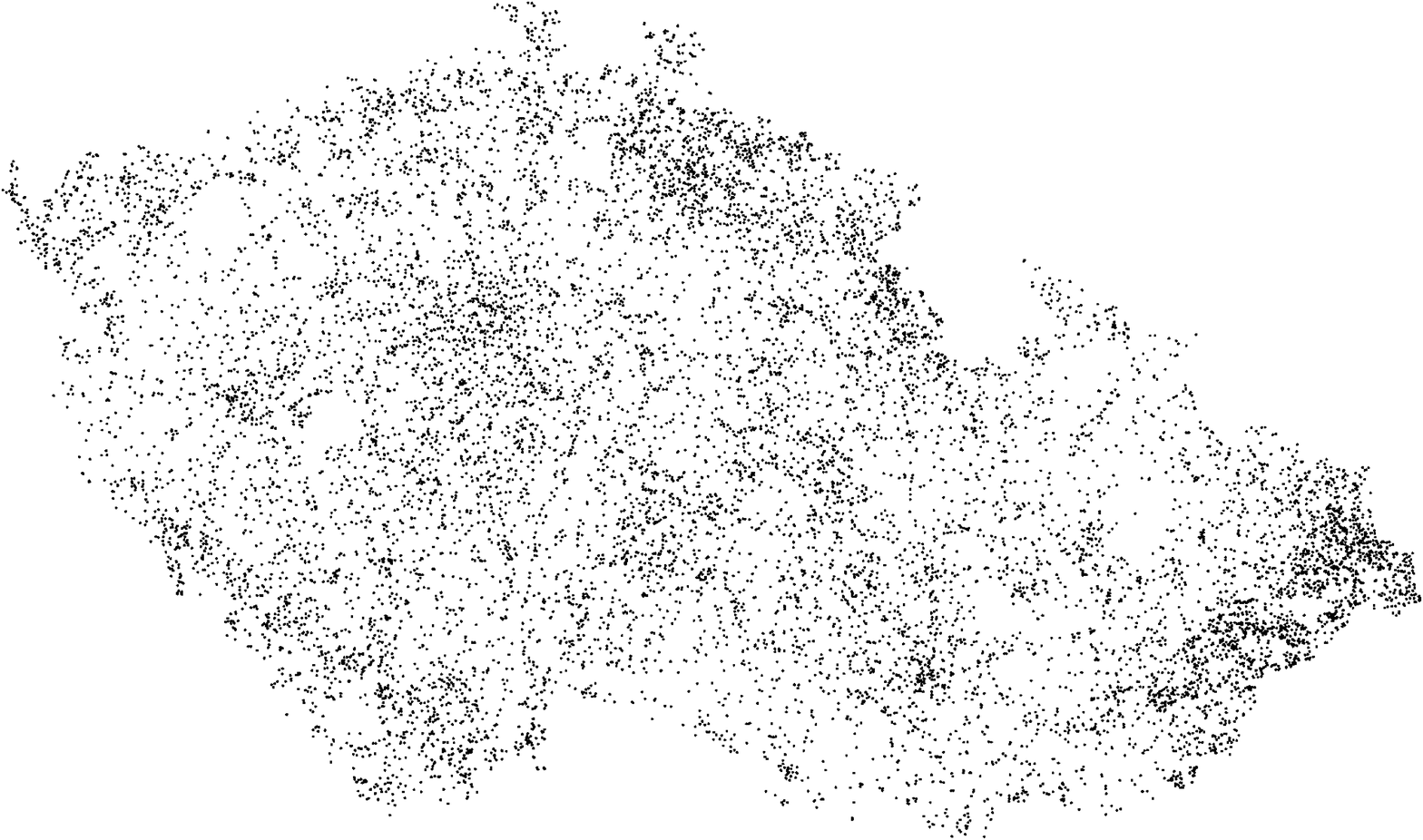}
  \caption{Sample 1\%}
	\label{fig:address200}
\endminipage\hfill
\end{figure*}

As may be seen from the figures, the first sample (see Figure \ref{fig:s1-5000}) preserves the cluster centers precisely. Especially the most dense areas are preserved. The points on the border of clusters and points with higher distance from the cluster centers are removed. The second sample (see Figure \ref{fig:s1-10000}) with higher radius show  a different aspect of the algorithm, because the points were removed mostly from cluster centers while the border points are preserved, but the most dense areas are still preserved. The last sample (see Figure \ref{fig:s1-20000}) with the highest radius shows that only the most dense cluster centers are still preserved and other clusters are replaced by the points from the whole cluster. Very interesting is that even points which are far from the center of any cluster and might be removed as a noise are preserved if they form a cluster with the neighborhood.

The second experiment uses a real world 2-dimensional dataset which contains all address points in the Czech Republic provided by the government \footnote{http://www.ruian.cz/ (in Czech)}. This data set contains 2,7640,903 address points with coordinates in S-JTSK coordination system (S-JTSK is a coord. system which was used since beginning of the 20th century in Czechoslovakia and the length unit is approximately one meter). The points are distributed more densely in the area of large cities, e.g. the most dense place is in the middle of the image where the capital city Prague is located, but very dense areas are also on the north and south, although the most populated area of the Czech Republic is on the east where Moravian-Silesian region is located. Similarly to the previous experiment we dicretize the distance with step 10. 
The summary of the data sampling with different radius are depicted in Table \ref{tab:czech} and for the visualization see Figures \ref{fig:address}-\ref{fig:address200}. 

\begin{table}[htbp]
  \centering
  \caption{Czech Map Dataset Sampling}
    \begin{tabular}{rrrr}
    \toprule
    \textbf{Log base} & \textbf{Radius} & \textbf{Points} & \textbf{Points \%} \\
    \midrule
    -   & -  & 2,740,903 & 100  \\
    1.3 & 50  & 206,603  & 8   \\
    1.3 & 100  & 55,641  & 2   \\
    1.3 & 200  & 21,965  & 1   \\
    \bottomrule
    \end{tabular}%
  \label{tab:czech}%
\end{table}%


As may be seen from the figures, visual comparison shows that the most dense parts are still dense and recognizable even when very large reduction is performed. The Figure \ref{fig:address100}, where only 2\% of points is preserved clearly show the largest cities in the Czech Republic and, moreover, shows that the east part of the republic has many densely populated places, and therefore, it is the most dense area preserved. 

\begin{remark}
We convert the problem of sampling the vector data to the problem of sampling networks as follows; every vector is a node that has an egde to every neighbor (a vector in its neighborhood) and the weight of the edge corresponds to the distance converted to similarity.
\end{remark}

\subsection{Algorithm complexity}
The complexity of the algorithm may be clearly extracted from the pseudo-code. If we suppose that the dataset D contains N objects and the average size of the neighborhood is $M$ then the complexity of the algorithm is $O(NM)$. Usually we may assume that the $M \ll N$ so the complexity is linear. This is done because of the locality of the algorithm. If we think deeper about the algorithm we see that the complexity is highly affected by the complexity of the neighborhood discovery. This is affected by the similarity and proximity functions, but if we suppose that these two functions follow the locality we may divide the problem into two types. First, when we deal with network data, we may suppose that the each node knows its neighbors because we usually have a list of edges for each node. When we deal with vector data the situation is different. In such dataset we know only the information about each vector itself but we have no information about its neighbors. So we must use some data structure which enables fast neighborhood exploration.  Many such structures were developed in the past, such as R-Tree and KD-Tree, and their variants or when the data has small dimension we may use Quadrant tree. These structures allow to find neighbors in constant or, in the worst case, logarithmic time so the efficiency of the algorithm is still very good.

\section{Conclusions}
In this paper we presented our 'work in progress' in the field of sampling large-scale data. The approach is based on finding the representatives in the input dataset. Measurement of representativeness is done by the analysis of local properties and nearest neighbors. We show in the experiments how we practically apply the method to the weighted networks and vector data. As the key features of the method we consider its applicability for weighted networks, natural scalability and generality.\\
There are several more tasks to solve in the future work. In particular, carrying out with experiments on large-scale data and comparing with other biased and unbiased methods. Since
every dataset requires a different setting it is necessary to make a deeper analysis of the dependencies between parameters of the presented method, processed dataset and expected representative sample.
We see the potential especially in the openness of the method to the definitions of similarity, proximity and representativeness. Given that all those functions may be non-symmetric, subject of experiments will also be the directed networks.

\section{Acknowledgments}
This work was supported by the European Regional Development Fund in the IT4Innovations Centre of Excellence project (CZ.1.05/1.1.00/02.0070), by the Development of human resources in research and development of latest soft computing methods and their application in practice project, reg. no. CZ.1.07/2.3.00/20.0072 funded by Operational Programme Education for Competitiveness, co-financed by ESF and state budget of the Czech Republic, and by SGS, VSB-Technical University of Ostrava, under the grant no. SP2014/110.

\bibliographystyle{abbrv}
\bibliography{SamplingACM}  
%

\balancecolumns
\end{document}